# A modal model for diffraction gratings

Mark P. Davidson
Spectel Research Corporation
807 Rorke Way
Palo Alto, CA 94303
mdavid@spectelresearch.com
www.spectelresearch.com

January 6, 2002

**ABSTRACT**

A description of an algorithm for a rather general modal grating calculation is presented. Arbitrary profiles, depth, and permittivity are allowed. Gratings built up from sub-gratings are allowed, as are coatings on the sidewalls of lines, and arbitrary complex structure. Conical angles and good conductors are supported.

## 1. INTRODUCTION

The modal approach to diffraction grating analysis has an extensive history partially represented by references [1-17]. In the present work a modal algorithm is presented which is exceptionally flexible and general. The method described allows sub-periods, sidewall coatings on grating lines, and in fact any complex geometry one may wish to consider so long as the geometry repeats itself periodically and has one axis of translational invariance. The grating can consist of any number of layers, any number of different materials in each layer, and any number of cells per layer. All materials can be dielectric or conducting, and the angles of incidence can be conical. The only limitations are of a practical numerical nature. Modal models have only recently been extended to handle more than two regions in a given horizontal layer [18]. The approach of [18] utilizes an equivalent electrical network technique in which each individual layer is described by a transmission-line unit. The present work does not use this method, but similar generality is achieved.

The main advantages of the modal method are that it is in theory more accurate than other methods and by careful study of the eigenfunctions it can aid in physical intuition. Another advantage is that the shapes of multilayer gratings can be modified without necessarily redoing all of the calculation. The main disadvantages are complexity and problems with numerical precision. In the current work it was found that for larger pitch values and for strong conductors it was necessary to work in 128 bit floating point or higher precision. The algorithm was implemented in several languages including Matlab, Mathematica, and C++ with double, double-double, and quad-double precision arithmetic. Mathematica's arbitrary precision arithmetic features were very attractive, but the run times were very long. A good compromise between speed and precision was found by running a preliminary calculation of the eigenvalues using double precision arithmetic, and then touching up these values by running a double-double or quad-double precision algorithm.

The application of grating simulation to metrology problems in the semiconductor industry has a long history [19-25]. The modal method has been applied in this context in [26-29], although the author was not aware of the earlier work at that time, and so the method was referred to as the Analytic Waveguide Method with the tradename Metrologia. Recent interest in grating simulation has increased because of the use of diffraction grating scatterometry for measuring linewidths and cross-sections in semiconductor manufacturing.

In the present paper only non-magnetic materials are considered as these are the most important in semiconductor manufacturing, although it seems straightforward theoretically at least to include magnetic materials.

## 2. MATHEMATICAL FORMULATION

Start with the fixed frequency complex form of Maxwell's equations which may be taken to be in appropriate units

$$\nabla \times \mathbf{E} - \frac{i\omega}{c}\mathbf{B} = 0 \tag{1}$$

$$\nabla \times \mathbf{B} + \frac{i\omega}{c}n^2\mathbf{E} = 0; \quad n^2 = \varepsilon + i\frac{\sigma}{\omega} \tag{2}$$

The object is to solve these equations given that a plane wave is incident on the grating of Fig. 1 from above

**Notation**:

$p$ = pitch or period

$\mathbf{k}^i$ = the incident wave vector before any scattering

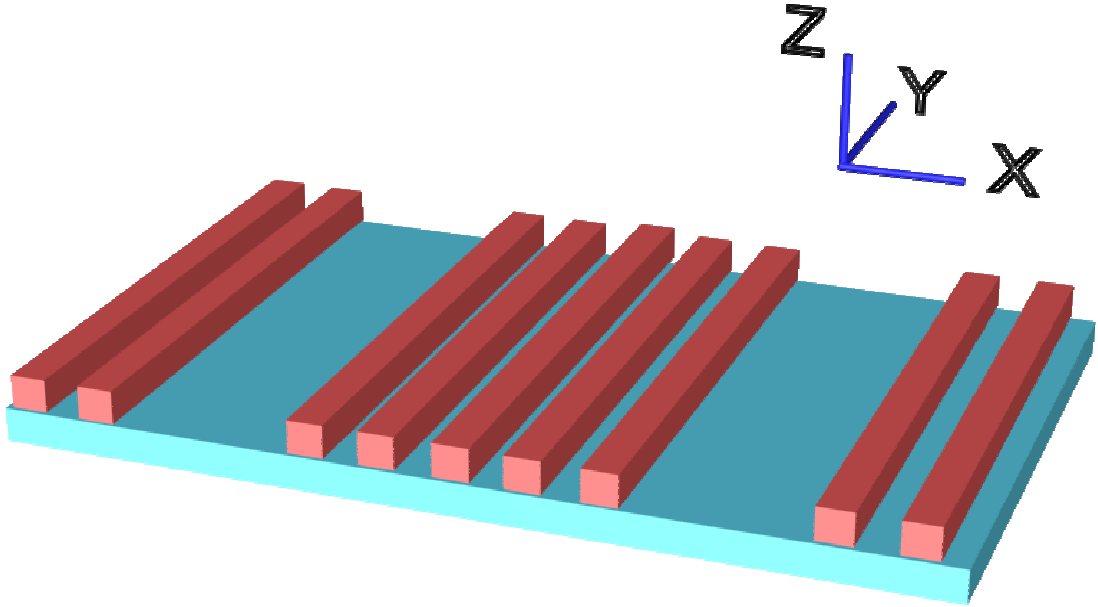

Figure 1

The scattering object is assumed to be periodic in the x direction and to be invariant in the y direction. The Floquet condition and the wave equation must be satisfied by all of the field components

$$F(x+p) = e^{ik_x^i p} F(x) \tag{3}$$

$$\left[\Delta + n(\mathbf{x})^2 k_0^2\right] F(\mathbf{x}) = 0; \quad k_0 = \frac{\omega}{c} = \frac{2\pi}{\lambda} \tag{4}$$

Distinguish two polarization states denoted by TE and TM. These are defined by the conditions that for TE the x component of the electric field vanishes, and for TM the x component of the magnetic field vanishes.

TE: $E_x = 0$     (5)

TM: $B_x = 0$     (6)

It can be shown that for a plane wave these states have polarization vectors which are orthogonal.

Barring exceptional circumstances a general field can be written as a superposition of a TE field and a TM field. The only case where this isn't possible is a plane wave travelling exactly in the x direction in some region. For such a plane, the x component of **E** and **B** are both zero because plane waves are transverse. Should one of the eigenmodes have behave this way, then by making a very small change to the wavelength or to the index of refraction one can make the z component of the wavevector **k** nonzero and thus eliminate this rare problem in practice.

Now consider an object consisting of vertical slab layers as in Figure 2. Both the superstrate and the substrate are assumed to be infinite in extent in the y and z directions.

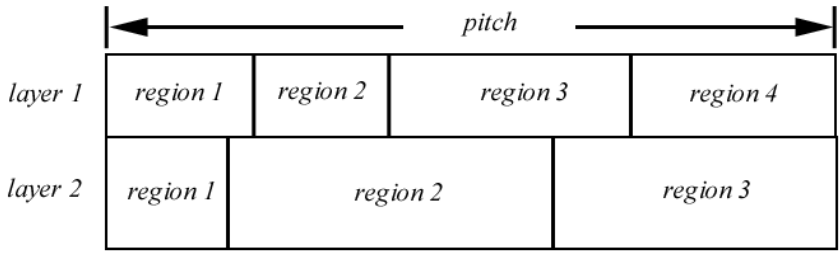

Figure 2

Only one period of the object is shown in the cross-section Figure 2, but the calculation assumes that the object repeats in the x direction. Each layer may have a different number of regions. For example, in figure 2 there are 4 regions in layer 1 and 3 regions in layer 2. The complex indexes of refraction can be different for each region. The method presented here works for any number of regions, for any number of layers, and for any number of different indexes of refraction.

Because of the invariance of the object in the y direction, one may factor out the y dependence of the field

$$F(x,y,z) = e^{ik_y \cdot y} F(x,z), \quad k_y = k_y^i \tag{7}$$

A solution within a given layer is found by the classical separation of variables method

$$F(x,z) = \sum_n A_n e^{ik_{z,n} \cdot z} \Phi_n(x) \tag{8}$$

The $k_{z,n}$ are the eigenvalues, they come in pairs with opposite sign, and the $\Phi$'s are the eigenfunctions. The condition which determines the eigenvalues is derived by imposing boundary conditions on $E_x$ and $B_x$.

The eigenvalues transform in a simply way as a function of $k_y$.

$$k_{z,n}(k_y) = \sqrt{k_{z,n}(0)^2 - k_y^2} \tag{9}$$

To see why this is so, imagine extending the layer under consideration to infinity above and below. This obviously won't affect the eigenvalues or eigenfunctions. But the infinite slab waveguide is now invariant under rotations about the x axis. A whole family of eigensolutions with different $k_y^i$ can be generated from one with $k_y^i = 0$ by exploiting this rotational invariance. Eq. (9) is simply a rotation about the x axis.

## 2.1 THE CONDITION DETERMINING THE EIGENVALUES

Let the primary field functions be $E_x$ and $B_x$. The other field quantities are derived from these using Maxwell's equations. In a given layer one starts with the boundary conditions for $E_x$, $B_x$ and their x derivatives. As one moves across the layer in the x direction, the following functions are continuous at the region boundaries (these are the vertical walls between regions in figure 2)

$$n^2 E_x, B_x, \frac{\partial E_x}{\partial x}, \text{ and } \frac{\partial B_x}{\partial x} \tag{10}$$

The eigencondition is found by continuing $E_x$ or $B_x$ an entire period in the x direction, and then imposing the Floquet condition on the field. For TM mode one continues $E_x$ and for TE mode $B_x$.

Consider a single eigenmode. In each region $l$ a given eigenmode has at most two plane waves whose $k_{x,l}$ values have opposite sign. This follows simply from the wave equation, the $k_{x,l}$ value being determined from the eigenvalue $k_{z,l}$ by the equation

$$k_{x,l}(x) = \pm \sqrt{n(x)^2 k_0^2 - k_{z,l}^2 - k_y^2} \tag{11}$$

There are only two values of $k_{x,l}$ because $k_y$ is unchanged by the scattering which follows from the translational invariance of the scattering object in the y direction. As a consequence $k_{x,l}$ is determined up to a sign by $k_z$ alone. In the general 3 dimensional case there would be an infinite number of $k_{x,l}$ values allowed in a given eigenmode thus complicating the problem considerably.

So in each region of a layer one can describe each eigenfunction in terms of a two-element complex vector representing the coefficients for the two allowed values of $k_x$. In each region one can write for each eigenfunction

$$\Phi_l(x) = a \cdot e^{ik_{x,l} \cdot (x - x_c)} + b \cdot e^{-ik_{x,l} \cdot (x - x_c)} \tag{12}$$

where $\Phi$ is equal to $B_x$ for the TE mode and to $E_x$ for the TM mode, and where $k_{x,l}$ denotes the positive square root in (11). The center of the region in question is $x_c$. The constants a and b will depend on the region.

So the eigenfield in each region is described by a two element vector (a,b) for each region. The boundary conditions relate the vectors in each region, and the Floquet condition imposes the final eigenvalue condition. Starting with the leftmost region, one can derive a matrix M which gives the a and b values 1 pitch to the right of the starting region.

$$\begin{pmatrix} a(p) \\ b(p) \end{pmatrix} = \mathbf{M} \cdot \begin{pmatrix} a(0) \\ b(0) \end{pmatrix} \tag{13}$$

where M is a product of matrices which couple neighboring regions

$$\mathbf{M} = \prod_{n=1}^{N} \mathbf{m_n} \tag{14}$$

where $m_n$ links the vector in region n+1 and region n. N is the number of regions in a single pitch or period. Figure 3 illustrates the situation

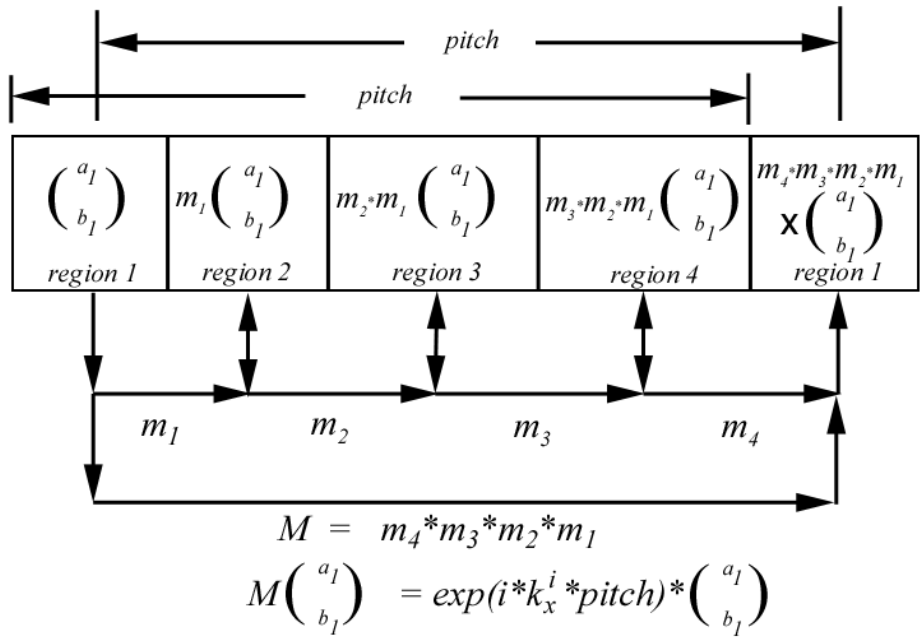

Figure 3

The matrices $m_i$ are different for TE and TM modes. After some algebra the boundary conditions reveal the following expressions for $m_i$ in the two modes:

$$\mathbf{m}^{TE} = \frac{1}{2} \left( \begin{bmatrix} (1+k_{xL}/k_{xR})* \\ \exp(ik_{xL}*WL)\exp(ik_{xR}*WR) \end{bmatrix} \begin{bmatrix} (1-k_{xL}/k_{xR})* \\ \exp(-ik_{xL}*WL)\exp(ik_{xR}*WR) \end{bmatrix} \\ \begin{bmatrix} (1-k_{xL}/k_{xR})* \\ \exp(ik_{xL}*WL)\exp(-ik_{xR}*WR) \end{bmatrix} \begin{bmatrix} (1+k_{xL}/k_{xR})* \\ \exp(-ik_{xL}*WL)\exp(-ik_{xR}*WR) \end{bmatrix} \right) \quad (15)$$

$$\mathbf{m}^{TM} = \frac{1}{2} \left( \begin{bmatrix} ((nL/nR)^2 + k_{xL}/k_{xR})* \\ \exp(ik_{xL}*WL)\exp(ik_{xR}*WR) \end{bmatrix} \begin{bmatrix} ((nL/nR)^2 - k_{xL}/k_{xR})* \\ \exp(-ik_{xL}*WL)\exp(ik_{xR}*WR) \end{bmatrix} \\ \begin{bmatrix} ((nL/nR)^2 - k_{xL}/k_{xR})* \\ \exp(ik_{xL}*WL)\exp(-ik_{xR}*WR) \end{bmatrix} \begin{bmatrix} ((nL/nR)^2 + k_{xL}/k_{xR})* \\ \exp(-ik_{xL}*WL)\exp(-ik_{xR}*WR) \end{bmatrix} \right) \quad (16)$$

Note that m couples two different regions, denoted by left (L) and right (R). The quantities in 15 and 16 have the following meanings:

$k_{xL}$ = x component of the k vector in the leftside region.
$k_{xR}$ = x component of the k vector in the rightside region.
WL = Half the width of the leftside region.
WR = Half the width of the rightside region.
NL = complex index of refraction in the leftside region
NR = complex index of refraction in the rightside region.

The eigenvalue condition is derived from Floquet's relation which states that

$$\mathbf{M}\begin{pmatrix} a \\ b \end{pmatrix} = \exp(ik_x^i p)\begin{pmatrix} a \\ b \end{pmatrix} \quad (17)$$

where p is the period or pitch. In order for a non-trivial solution to this equation, the following determinant must vanish:

$$\det(\mathbf{M} - \exp(ik_x^i p)\mathbf{1}) = 0 \quad (18)$$

It follows easily from the definitions of **M** that

$$\det(\mathbf{M}) = 1 \quad (19)$$

Since **M** is dimension 2, a more useful form for the eigenvalue relation can be obtained. The problem with (18) is that the elements of **M** can be very large in some cases. For example when good conductors are being modeled or when $k_x$ is imaginary in some of the regions making up the layer. The following derivation gives a more useful form which does not require as much numerical precision. Since **M** has nonzero determinant, it can be diagonalized with a similarity transformation without changing (18). In terms of the eigenvalues of **M**, (18) becomes

$$(\lambda_1 - \varphi)(\lambda_2 - \varphi) = 0 \tag{20}$$

where $\lambda_1$ and $\lambda_2$ are the eigenvalues of **M** and

$$\varphi = \exp(ik_x^i p) \tag{21}$$

$$\lambda_1 \lambda_2 = 1 \tag{22}$$

And therefore (20) becomes

$$1 - (\lambda_1 + \lambda_2)\varphi - \varphi^2 = 0 \tag{23}$$

But the trace of **M** is invariant under a similarity transformation also, and therefore one has

$$1 - \text{trace}(\mathbf{M})\varphi - \varphi^2 = 0 \tag{24}$$

The eigenvalue condition expressed in the form (24) is superior to (18) for numerical reasons because the trace is first order in the elements of **M** and therefore when the elements of **M** are large, the precision required is less for (24) than for (18). The eigenvalue condition is satisfied by finding those $k_x$ values which satisfy (24). Once a $k_x$ is found then $k_z$ follows from the wave equation.

## 2.2 MORPHING TO FIND NEW EIGENVALUES FROM KNOWN ONES

There is no general algorithm for finding all the complex zeros of an analytic function in the complex plane. Solving the eigenvalue problem is therefore nontrivial. Moreover, the presence of exponentials in eqns. 15 and 16 make numerical precision a problem in some cases. In [9, 30] an approach was presented which was based on contour integral theorems for analytic functions. The approach used here is a generalization of the Tayeb-Petit method [11] which was also used successfully in [12, 13, 26-29]. In order to solve the general eigenvalue problem one starts with a known solution and then varies the parameters of the problem slowly and continuously to change the object from the known one to the desired case, all the while tracking the continuous motion of the eigenvalues. A fictitious time variable t is introduced, with all parameters of the problem linear functions of this variable as follows:

$$P(t) = P(0) + t*(P(1)-P(0)) \tag{25}$$

The parameter t varies from 0 to 1. A convenient starting case for which the solutions are known is simply a homogeneous layer for which there is only one region. In the homogeneous case, it is easy to show that the allowed eigenvalues for $k_x$ are

$$k_x = k_x^i + N\frac{2\pi}{p}, \quad N = \text{integer} \tag{26}$$

Some of these eigenvalues will have the same $k_z$ (ie. will be degnerate) if $k_x$ is any multiple of

$$k_x^i = \frac{\pi}{p} \tag{27}$$

These degenerate cases cause a problem for the morphing algorithm, and so one choose a starting value for $k_x^i$ which is not a multiple of (27). The determinant to be zeroed can be thought of as a function of $k_x$ in the leftmost region of the layer and of t. As t varies from 0 to 1 a given eigenvalue $k_x$ will move on some trajectory. Let

$$F(k_x, t) = 1 - \text{trace}(\mathbf{M})\varphi - \varphi^2 = 0 \tag{28}$$

From this equation one can derive the trajectory equation:

$$\frac{dk_x}{dt} = \frac{-\frac{\partial F}{\partial t}\big|_{kx}}{\frac{\partial F}{\partial k_x}\big|_t} \tag{29}$$

A procedure that works well for solving this equation is to use (29) computed numerically to integrate a certain number of time steps, and then to periodically interrupt the integration in order to refine the accuracy of $k_x$ by using a Newton's method to minimize F as a function of $k_x$ for fixed t. It is important to choose the time steps to be small enough so that nearly degenerate eigenvalues can be resolved numerically. Degenerate cases are hard to avoid though, and some means for removing degenerate eigenvalues is very helpful.

## 2.3 FIXING DEGENERACY PROBLEMS

F in (28) will always have an infinite number of zeros as a function of $k_x$. No theorems are known which prove that degenerate eigenvalues cannot occur in the general case. Degeneracy certainly occurs for homogeneous layers when $k_x^i$ is a multiple of (27). Certainly cases can arise when eigenvalues can come close together, and this can cause numerical difficulty. It appears that true degeneracy does occur for sub periodic gratings, like the one pictured in Figure 1 if the lines are all identical and equally spaced in groups. The degeneracy can be split in this case by choosing the lines and spaces to be very slightly different. One way to deal with near collisions of eigenvalues is to choose a very small time step when integrating (29). This can be quite time consuming in the general case however. It has proved more efficient to allow some degeneracy and then remove it using the following technique.

Suppose that two nearby eigenvalues have been confused by the integration procedure (29) together with Newton's method. So two eigenvalues are incorrectly thought to be the same. It is to be expected that actually only one of the eigenvalues is the value that has been calculated for both, and the other one is different and somewhere nearby. Then F will have a simple zero at the point found. And therefore the following function

$$F_1 = F/(k_x - k_{x_1}) \tag{30}$$

will have a zero at the eigenvalue which was missed, but not at $kx_1$ anymore. If there are several zeros which have been identified in the neighborhood of $kx_1$, then these can be divided out also to make the chance of finding the correct zero better. In this case one uses a function of the form

$$F_1 = \frac{F}{\prod_n (k_x - k_{x_n})} \tag{31}$$

This method is not completely foolproof, but if there are only a few degeneracies it will often remove them. If not, then the other alternative is to choose the time step to be smaller.

## 2.4 DETERMINING THE ELECTRIC AND MAGNETIC FIELD VECTORS FOR A GIVEN EIGENFUNCTION

In TE mode work with $B_x$. Normalize the plane wave in each region such that $B_x = 1$.

$$TE: \quad \mathbf{E} = \mathbf{k} \times \hat{\mathbf{x}}, \quad \mathbf{B} = \frac{1}{k_0} \mathbf{k} \times \mathbf{E}, \text{ normalize this vector so that } B_x = 1 \tag{32}$$

In TM mode work with $E_x$, and normalize the plane waves in each region so that $E_x = 1$.

$$TM: \quad \mathbf{B} = \mathbf{k} \times \hat{\mathbf{x}}, \quad \mathbf{E} = -\frac{1}{n^2 k_0} \mathbf{k} \times \mathbf{B}, \text{ normalize this vector so that } E_x = 1 \tag{33}$$

The amplitudes or functional forms for $E_x$ and $B_x$ as functions of x are found after solving the eigenvalue condition. Then these eigenfunctions are multiplied by the above vectors in each region and for each plane wave making up the eigenfunction to get the full complement of fields.

## 2.5 DETERMINING THE EIGENFUNCTIONS

Finding the eigenfunctions amounts to finding the two numbers a and b in each region and using equation 12 to get the functional behavior. One starts by finding this vector in the leftmost region of a layer. The eigenfunction is determined up to a muliplicative constant, and so without loss of generality (provided that $M_{01}$ is not zero) one can set a = 1 in the leftmost region.

$$M \begin{pmatrix} a_1 \\ b_1 \end{pmatrix} = \varphi \begin{pmatrix} a_1 \\ b_1 \end{pmatrix} \Rightarrow b_1 = -\frac{(M_{00} - \varphi)}{M_{01}} a_1 \tag{34}$$

So with this relation the eigenfunction in region 1 is determined. If it so happens that $M_{01}$ is zero, then either only a or b can be nonzero, or if $\varphi=1$ then a and b can both be non zero but are independent of one another. This happens in homogeneous layers. To find the vector in region 2 simply multiply this vector by $m_1$

$$\begin{pmatrix} a_2 \\ b_2 \end{pmatrix} = \mathbf{m}_1 \begin{pmatrix} a_1 \\ b_1 \end{pmatrix}, \begin{pmatrix} a_3 \\ b3 \end{pmatrix} = \mathbf{m}_2 \begin{pmatrix} a_2 \\ b_2 \end{pmatrix}, etc. \tag{35}$$

where of course one must use the appropriate form (TE or TM) for M and $m_i$. In this way the eigenfunctions are found for all the regions, and then the field quantities are found by multiplying these eigenfunctions by the basis vectors derived in section 2.4.

## 2.6 ADJOINT EIGENFUNCTIONS AND THE HILBERT SPACE NORM

As discussed by previous authors [8, 12], one must use the theory of non self-adjoint operators in order to proceed with the scattering calculation for cases where the index of refraction is complex. A different Hilbert space norm is used here from all of the previous papers on modal models and it involves only the x components of the fields. This is the most natural choice for inner product since for the TE and TM modes, one of the basis fields ($E_x$ or $B_x$) always vanishes. Using Maxwell's equation, one can derive the following equations for the x components of the fields in a layer.

$$\left[ n(x)^2 k_0^2 + \frac{\partial^2}{\partial x^2} + \frac{\partial^2}{\partial y^2} + \frac{\partial^2}{\partial z^2} \right] B_x = 0 \tag{36}$$

$$\left[ n(x)^2 k_0^2 + \frac{\partial}{\partial x} \frac{1}{n(x)^2} \frac{\partial}{\partial x} n(x)^2 + \frac{\partial^2}{\partial y^2} + \frac{\partial^2}{\partial z^2} \right] E_x = 0 \tag{37}$$

A suitable inner product is

$$\langle u,v \rangle = \int_0^{pitch} \left[ B^u_x{}^\dagger B^v_x + n(x)^2 E^u_x{}^\dagger E^v_x \right] dx \tag{38}$$

This equation expresses a major difference between the current algorithm and all the previous modal algorithms which work with the field components parallel to the lines. The advantage of the inner product (38) is that it is immediately obvious that the TE and TM eigenfunctions are orthogonal with respect to this norm since for TE mode $E_x$ is zero and for TM mode $B_x$ is zero. Moreover it is simpler to calculate than the norm used for example in [12]. It is easy to show that the following differential operators are self adjoint with respect to this inner product provided that $n(x)$ is real

$$L_B = n(x)^2 k_0^2 + \frac{\partial^2}{\partial x^2} \tag{39}$$

$$L_E = n(x)^2 k_0^2 + \frac{\partial}{\partial x} \frac{1}{n(x)^2} \frac{\partial}{\partial x} n(x)^2 \tag{40}$$

When $n(x)$ is not purely real, the operators $L_B$ and $L_E$ are no longer self adjoint. To be able to proceed with a modal expansion for the total field, the adjoint operators and eigenfunctions must be considered. The adjoint operators are

$$L_B^\dagger = \left(n(x)^*\right)^2 k_0^2 + \frac{\partial^2}{\partial x^2} \tag{41}$$

$$L_E^\dagger = \left(n(x)^*\right)^2 k_0^2 + \frac{\partial}{\partial x} \frac{1}{\left(n(x)^*\right)^2} \frac{\partial}{\partial x} \left(n(x)^*\right)^2 \tag{42}$$

The adjoint eigenfunctions for $B_x$ (TE mode) satisfies the equation

$$L_B^\dagger B_x = (k_y^2 + k_z^{*2}) B_x \tag{43}$$

along with the Floquet condition 17. The adjoint eigenfunction for $E_x$ (TM mode) satisfies

$$L_E^\dagger E_x = (k_y^2 + k_z^{*2}) E_x \tag{44}$$

again with the Floquet condition.

The adjoint eigenfunctions are generated by using the matrix functions (15) and (16) but using the conjugate values for the k's and n's.

The eigensolutions come in pairs corresponding to opposite signs of $k_z$. The fields within a given layer can be expanded in the eigenfunctions of that layer

$$\mathbf{F} = \sum_n b_n \mathbf{F_n^{TE}}(x) B_x^n(x) e^{ik_y y + ik_{z,n} z} + \sum_n e_n \mathbf{F_n^{TM}}(x) E_x^n(x) e^{ik_y y + ik_{z,n} z} \qquad (45)$$

In this expression, $\mathbf{F}$ is a 6 dimensional vector $(E_x, E_y, E_z, B_x, B_y, B_z)$ containing both fields. The quantities $\mathbf{F_n^{TE}}(x)$ and $\mathbf{F_n^{TM}}(x)$ are the (6d) polarization basis vectors for TE and TM modes calculated in section 2.4. They depend in a stepwise way on the region which is determined by the value of x. The functions $E_x^n(x)$ and $B_x^n(x)$ are the normalized eigenfunctions for $E_x$ and $B_x$ of section 2.5. The complex numbers $b_n$ and $e_n$ are the amplitudes of the nth TE and TM mode respectively. The index n refers to which mode is being considered. The modes come in pairs which have opposite signs for $k_z$.

The eigenfields for the two signs of $k_z$ for a given pair are not orthogonal. However, they can be distinguised by considering z derivatives of $\mathbf{F}$. For such a pair the functions $B_x^n$ and $E_x^n$ are the same, but the vector fields $\mathbf{F_n^{TE}}$ differ. A projection operation can be defined which projects out the positive $k_z$ and negative $k_z$ contributions as follows

$$\langle \varphi_n | \mathbf{F}_\pm \rangle = \frac{ik_{z,n} \langle \varphi_n | \mathbf{F} \rangle \pm \langle \varphi_n | \partial \mathbf{F} / dz \rangle}{2ik_{z,n}} \qquad (46)$$

In general $k_z$ can be complex, and so deciding whether to group a given $k_z$ with the plus or minus set is possibly ambiguous. The only layers where the choice makes a difference is in the superstrate and the substrate. In the substrate one makes a distinction between causally allowed waves (downward traveling or exponentially decaying) and forbidden waves (upward traveling or exponentially growing). In the substrate therefore one can group all forbidden waves together and all allowed waves together and label them + and - respectively.

## 2.7 BOUNDARY CONDITIONS AT LAYER INTERFACES

All components of B are continuous at the boundary. The tangential components of E are continuous and $n^2 E_z$ is continuous. The object is to be able to calculate the fields just below the layer interface given the fields just above the layer interface. The following relations follow from the basic boundary conditions together with Maxwell's equations

$$E_x \big|_{lower} = E_x \big|_{upper} \qquad (47)$$

$$\frac{\partial E_x}{\partial z} \bigg|_{lower} = \frac{n_{upper}^2}{n_{lower}^2} \frac{\partial E_x}{\partial z} \bigg|_{upper} - i\omega \left( \frac{n_{upper}^2}{n_{lower}^2} - 1 \right) B_y \big|_{upper} \qquad (48)$$

$$B_x\big|_{lower} = B_x\big|_{upper} \tag{49}$$

$$\frac{\partial B_x}{\partial z}\bigg|_{lower} = \left(1 - \frac{n_{lower}^2}{n_{upper}^2}\right)\frac{\partial B_z}{\partial x}\bigg|_{upper} + \frac{n_{lower}^2}{n_{upper}^2}\frac{\partial B_x}{\partial z}\bigg|_{upper} \tag{50}$$

$$\frac{\partial B_y}{\partial z}\bigg|_{lower} = \frac{n_{lower}^2}{n_{upper}^2}\frac{\partial B_y}{\partial z}\bigg|_{upper} + ik_{yin}(1 - \frac{n_{lower}^2}{n_{upper}^2})B_z\bigg|_{upper} \tag{51}$$

## 2.8 LAYER TO LAYER OVERLAP INTEGRALS

The scattering calculation boils down to calculating the overlap integral between a single eigenfield in the upper layer and the eigenfields in the lower layer. Once this matrix is known for each layer interface it is easy to solve the scattering problem.

Start with a single eigenmode in the upper layer. Apply (47-51) to the eigenfield to calculate the fields in the lower layer. The continued fields will no long be a single eigenmode, but will be a superposition of the eigenmodes in that layer. The lower layer will in general contain upward traveling and downward traveling waves (both signs of $k_z$). The components of the various eigenmodes in the lower layer are calculated using (46).

There is a complication to this procedure. The following class of integrals must be calculated:

$$I = \int n_{lower}^2 \varphi_n^\dagger \frac{dE_z}{dx}\bigg|_{lower} dx \tag{52}$$

The problem is that $\frac{dE_z}{dx}$ has singularities at the region boundaries which complicate the integral. The following artifice facilitates the calculation of (52). Notice that

$$\frac{dE_x}{dz} - \frac{dE_z}{dx} = ik_0 B_y \tag{53}$$

But $B_y$ is continuous at all boundaries. Substituting into (52) one obtains

$$I = \int n_{lower}^2 \varphi_n^\dagger (\frac{dE_z}{dx} + ik_0 B_z)\bigg|_{lower} dx \tag{54}$$

Now comes the trick. Integrate the first term by parts to eliminate the singular behavior.

$$I = \int n_{lower}^2 \varphi_n^\dagger ik_0 B_y dx - \int \frac{dn_{lower}^2 \varphi_n^\dagger}{dx} E_z\big|_{lower} dx \tag{55}$$

In this form the integral is easy to evaluate numerically. Without this technique it is quite difficult to get the algorithm to work properly using the norm (38).

## 2.9 SOLUTION TO THE SCATTERING PROBLEM

The Field in a given layer is described by a state vector of amplitudes for the various eigenmodes. Once the state vector is known in a given layer, it can be inferred for all other layers. Upward traveling and downward traveling modes with opposite signs for $k_z$ may be considered as distinct modes for the purpose of this discussion. Practical considerations require that the infinite series be approximated by a finite size vector of dimension N. For simplicity let's assume that the dimension at each layer is the same, although this isn't strictly necessary. Then the overlap matrix is always an N by N matrix. What is desired is a matrix - called the T matrix - which couples a state vector in the superstrate with a state vector in the substrate. It can be calculated from the following formula:

$$\mathbf{T} = O_{N_L+1,N_L} \cdot \prod_{j=1}^{N_L} \exp(ik_z^j h_j) O_{j,j-1} \tag{56}$$

where O is the N by N overlap matrix and vector array notation is used for the exponentiation and where $k_z$ is an N dimensional vector signifying the $k_z$ values for the N eigenmodes and $h_j$ is the vertical thickness of the layer j. There are $N_L$ layers between the superstrate and the substrate.

In order to solve the scattering problem one must break T up into 4 submatrices corresponding to coupling between upward and downward traveling waves in the substrate and in the superstrate. Let an up arrow denote an upward traveling wave and a down arrow a downward traveling wave. An exponentially decaying wave which decays in the upward direction will also have an up arrow and one that decays in the downward direction will have a down arrow. Consistency requires that for a complex index of refraction in the substrate that the real and imaginary part of $k_z$ have the same sign. This turns out always to be the case in our experience with the software. So the T matrix may be written:

$$\mathbf{T} = \begin{bmatrix} T_{\uparrow\uparrow} & T_{\uparrow\downarrow} \\ T_{\downarrow\uparrow} & T_{\downarrow\downarrow} \end{bmatrix} \tag{57}$$

The dimension of each sub matrix is N/2 by N/2. The scattering equation is

$$\begin{pmatrix} \text{substrate} \uparrow \\ \text{substrate} \downarrow \end{pmatrix} = \begin{bmatrix} T_{\uparrow\uparrow} & T_{\uparrow\downarrow} \\ T_{\downarrow\uparrow} & T_{\downarrow\downarrow} \end{bmatrix} \begin{pmatrix} \text{superstrate} \uparrow \\ \text{superstrate} \downarrow \end{pmatrix} \quad (58)$$

but causality requires *substrate* $\uparrow = 0$ and therefore one may derive that

$$T_{\uparrow\uparrow} \text{superstrate} \uparrow + T_{\uparrow\downarrow} \text{superstrate} \downarrow = 0 \quad (59)$$

Solving for the unknown reflected wave one obtains

$$\text{superstrate} \uparrow = -T_{\uparrow\uparrow}^{-1} T_{\uparrow\downarrow} \text{superstrate} \downarrow \quad (60)$$

So the reflecting S matrix is

$$S_{ref} = -T_{\uparrow\uparrow}^{-1} T_{\uparrow\downarrow} \quad (61)$$

Likewise the transmission S matrix is

$$S_{tran} = T_{\downarrow\uparrow} S_{ref} + T_{\downarrow\downarrow} \quad (62)$$

The result of one calculation gives a multitude of diffraction orders corresponding to the incoming and outgoing $k_x$ value taking values of $k_x^i$ plus a integer multiples of $2\pi/p$

## 2.10 CALCULATION OF DIFFRACTION EFFICIENCIES

Whenever $k_y$ is zero, the TE and TM modes decouple. That is, if the incoming wave is purely TE (TM) then the outgoing wave is purely TE (TM) also. When $k_y$ is not zero the two polarization modes are coupled. Let us consider a purely dielectric superstrate for reflection, and a purely dielectric superstrate and substrate for transmission. Also or transmission, let the superstrate and substrate be the same material. The formulas for the diffraction efficiency matrix are

For reflection:

$$D_{ref}[out,in] = \left| \frac{k_{zout}^2 + k_y^2}{k_{zin}^2 + k_y^2} \frac{k_{zin}}{k_{zout}} \right| \left| S_{ref}[out,in] \right|^2 \quad (63)$$

For transmission:

$$D_{tran}[out,in] = \left| \frac{k_{zout}^2 + k_y^2}{k_{zin}^2 + k_y^2} \frac{k_{zin}}{k_{zout}} \right| \left| S_{tran}[out,in] \right|^2 \quad (64)$$

The reciprocity property of Maxwell's equations requires that in many cases D is a symmetrical matrix.  This is more usually the case for reflection mode, but is also true in transmission if the scattering object is invariant under a parity transformation in the z coordinate.

**2.11  DISCUSSION OF NUMERICAL RESULTS**

The algorithm has been coded in C++ in three different versions using double precision, double-double precision and quad-double precision arithmetic [31].  It was found that double-double precision arithmetic was required for all but the simplest of problems.  Comparisons with the results tabulated in Table 1 of [29] are shown in figure 4.   The IESMP and Metrologia models are described in [29].   Only TM results are shown since the TE case is not particularly challenging, and the agreement is several orders of magnitude better for TE than for TM.  The reciprocity error for TE is also a much smaller value of 1.65e-07 (averaged over the range of heights in figure 4).

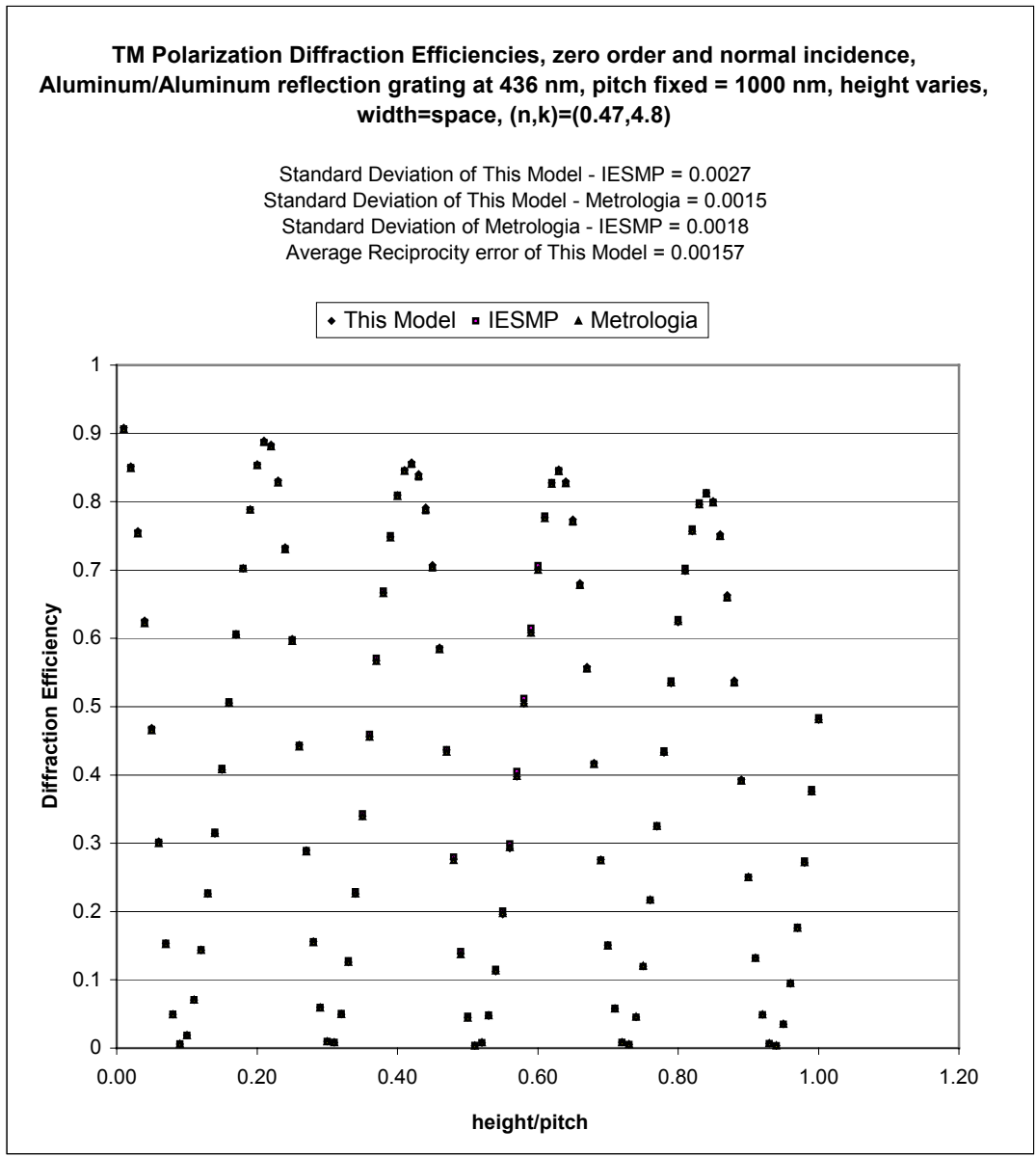

Figure 4

Figure 5 shows a comparison of calculations of the glass transmission grating of Figure 7 in [29].

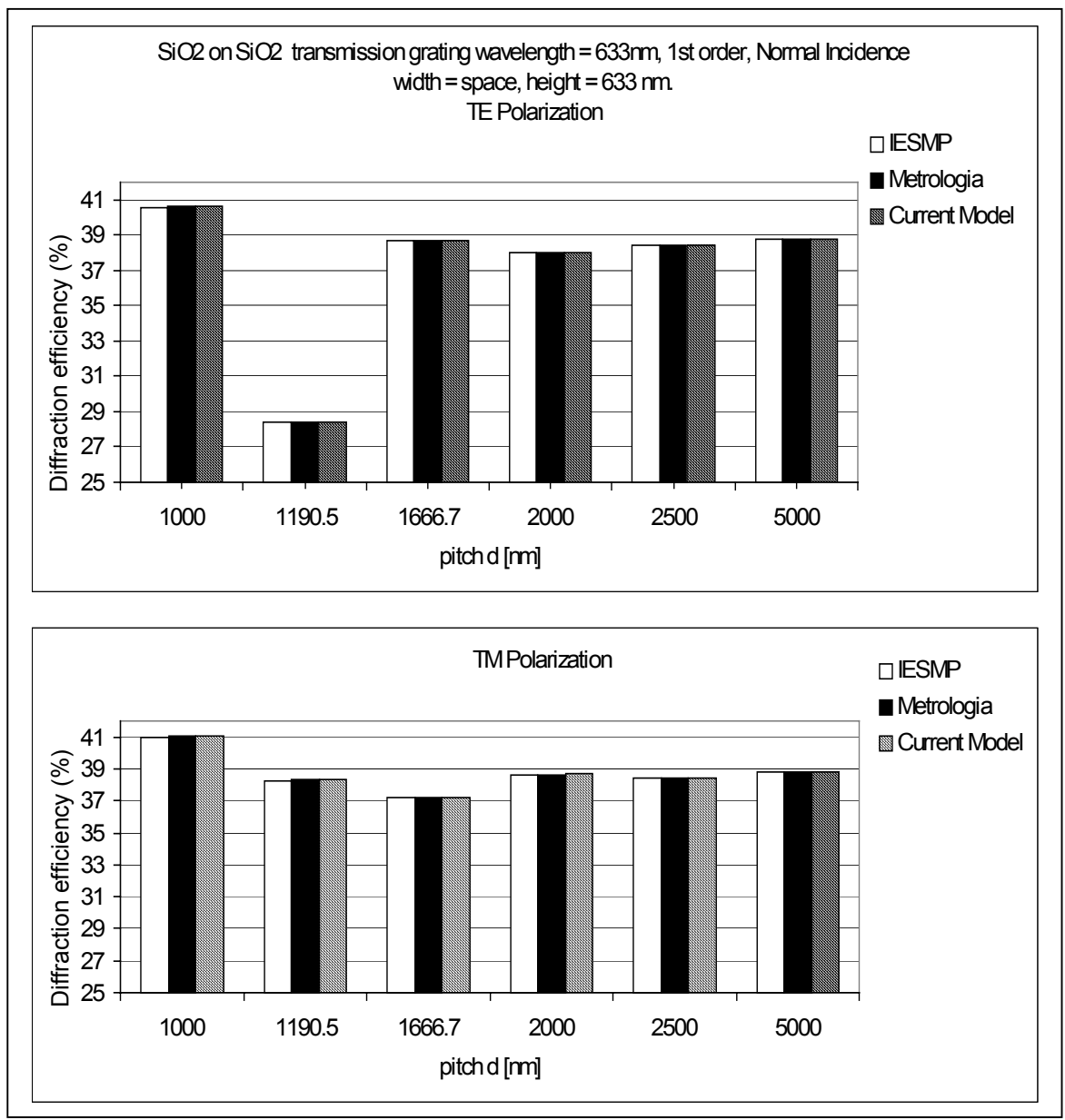

Figure 5

In Figures 6 and 7 a comparison is made between the current model and a commercial code which will be described here as Model X where the incident angles varies from 0 to 75 degrees.

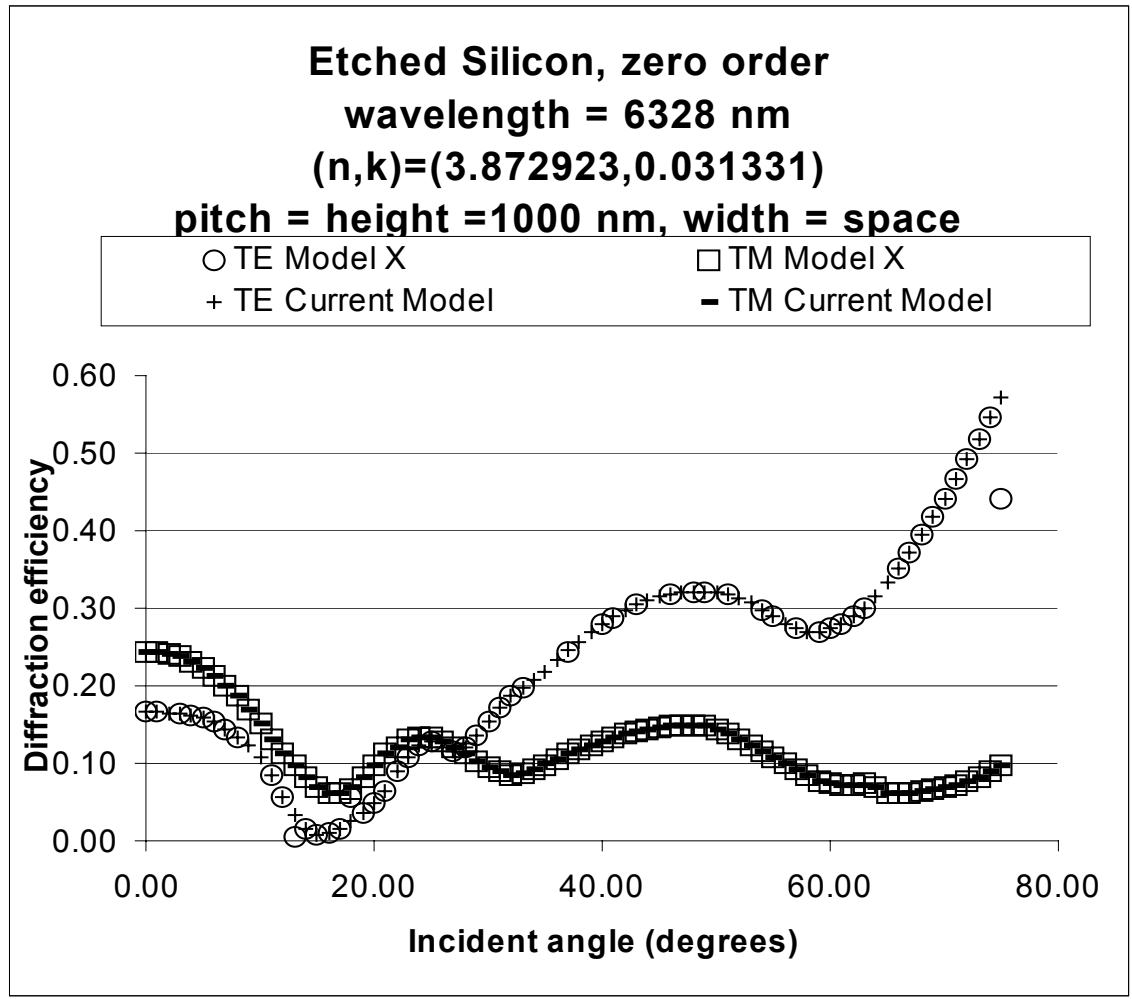

Figure 6

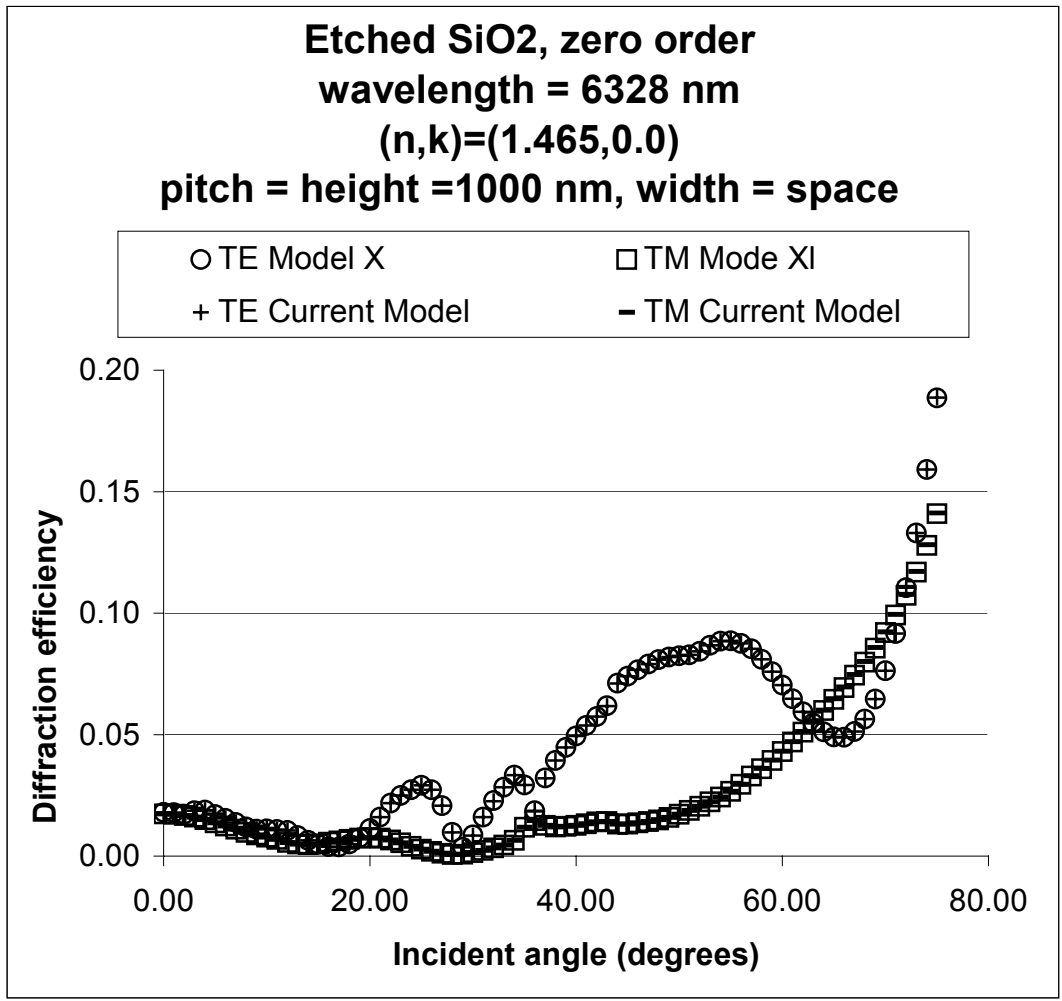

Figure 7

Table 1 shows a comparison between the current model and the Metrologia model for a mulitlayered structure.

**Description of Object**
    Normal Incidence Illumination
    Wavelength = 6328 nm; pitch = 1000 nm; 5 layer object, etched SiO2
    (n,k)=(1.465,0.0)
    Layer #1: width of SiO2 = 300 nm, height = 100 nm
    Layer #2: width of SiO2 = 350 nm, height = 100 nm
    Layer #3: width of SiO2 = 400 nm, height = 100 nm
    Layer #4: width of SiO2 = 450 nm, height = 100 nm
    Layer #5: width of SiO2 = 500 nm, height = 100 nm

| Diffraction Order | TE Metrologia | TE Current Model |
|---|---|---|
| 0 | 0.0008208 | 0.0008125 |
| 1 | 0.0131281 | 0.0131348 |

| | TM Metrologia | TM Current Model |
|---|---|---|
| 0 | 0.0004078 | 0.0004104 |
| 1 | 0.0106176 | 0.0106099 |

Table 1

Reciprocity of the solutions was calculated when relevent, and used as a sanity check on numerical solutions. Large pitches, large numbers of regions, and highly conductive materials all posed challenges for the algorithm. The principle problems encountered were:

1. Eigenvalue degeneracies that couldn't be removed by the techniques in 2.3. This problem can be addressed by choosing larger number of morphing steps and by adding small random dimensions to repeating lines in a sub-periodic array.

2. Insufficient numerical precision in calculating the eigenfunctions. This can be addressed by moving to quad or higher precision.

Despite these problems, many non-trivial problems showed good reciprocity. For example the data in table 2 shows results for a 5 line sub periodic structure similar to the one shown in Figure 1. The results are presented without a comparison since none was available.

> **Description of Object**
> Wavelength = 248nm, pitch=3000nm, materials: $SiO_2$ over Si.
> For $SiO_2$ (n,k)=(1.508, 0.0)
> For Si    (n,k)=(1.68, 3.58)
> Linewidths = 100nm, spaces = 100nm, height=100nm
> Target consists of sets of 5 equally spaced lines separated by 2100 nm gaps.
> Calculation results for reflection at **normal incidence.**
> Calculated TE reciprocity (worst off diagonal difference of the diffraction efficiency matrix) = 1.3e-4
> Calculated TM reciprocity= 1.0e-5
>
> Diffraction Efficiencies
>
> | Order | TE Result | TM Result |
> |---|---|---|
> | 0 | 0.25843067 | 0.46136406 |
> | 1 | 0.13456641 | 0.05222133 |
> | 2 | 0.034869257 | 0.013221011 |
> | 3 | 6.5723882e-005 | 6.0562931e-005 |
> | 4 | 0.0082766007 | 0.0039986665 |
> | 5 | 0.0048356392 | 0.0027805878 |
> | 6 | 0.00023264319 | 6.4767116e-006 |
> | 7 | 0.0048731602 | 0.0015477937 |
> | 8 | 0.0022670613 | 0.0015721377 |
> | 9 | 0.00098247656 | 5.5515111e-005 |
> | 10 | 0.0055509911 | 0.0010660416 |
> | 11 | 0.00065724551 | 0.002499388 |
> | 12 | 0.0050951529 | 0.0011042975 |

Table 2

## 3.0 CONCLUSION

The modal method can be extended in the way presented here to quite general grating structures. The main difficulty with the implementation is the apparent need to use higher precision arithmetic for larger structures or for strong conductors. Nevertheless, the technique offers promise as a tool for complex grating analysis. The model could benefit from an improved method for calculating the eigenvalues, perhaps replacing some of the numerical estimates of derivatives with analytic expressions.

**ACKNOWLEDGEMENTS**

The author acknowledges valuable discussions with Yiping Xu, Ron Herschel, Gordon Kino, Bob Larrabbee, Rick Silver, and Egon Marx during the long course of this work. He also acknowledges the helpful and diligent suggestions made by the reviewers and implemented in the final version.